\documentclass[12pt]{article}
\usepackage{graphicx}
\usepackage{amsmath} 
\usepackage{caption}
\usepackage{subcaption}
\usepackage{amssymb}
\usepackage{float}

\begin{document}
\DeclareGraphicsExtensions{.jpg,.pdf,.mps,.png}
\begin{sf}

\def\bp{\bar p}

\begin{center}

    {\bf BACKGAMMON: A Scheme for Producing High Intensity Muon  Beams for Future Colliders and Other Applications}\\

\vskip .15in

   {\bf Armen Apyan$^a$, Peter Delfyett$^b$, Paul Guèye$^c$, Letrell Harris$^d$ \\Sokhna Bineta Lo Amar$^c$,  Sekazi K. Mtingwa$^{e,\ast}$}\\

\vskip .20in

\vskip .20in
\end{center}

\noindent $^a$ A. Alikhanyan National Laboratory, 0036 Yerevan, Armenia \\
$^b$ University of Central Florida, CREOL, The College of Optics and Photonics, 4304 Scorpius Street, Orlando, Florida, 32816, USA  \\
$^c$ Michigan State University, Facility for Rare Isotope Beams, 640 South Shaw Lane, East Lansing, Michigan, 48824, USA \\
$^d$ University of Hawaii at Mānoa, Department of Physics and Astronomy, 2505 Correa Road, Honolulu, Hawaii, 96822 , USA \\
$^e$ TriSEED Consultants, LLC, Hillsborough, North Carolina, 27278, USA \\  

\vskip .05in
\noindent $^{\ast}$ Corresponding author.\\
Email addresses: aapyan@gmail.com (A. Apyan), delfyett@creol.ucf.edu (P. Delfyett), Gueye@frib.msu.edu (P. Guèye), lkharris@hawaii.edu (L. Harris), sokhnabinetalo.amar@ucad.edu.sn (S. Lo Amar), sekazi.mtingwa@gmail.com (S. Mtingwa).

\newpage

\begin{abstract}

We present a scheme for producing intense $\mu^+$ and $\mu^-$ lepton beams that could be utilized in a future muon-muon or muon-ion collider, as well as for other applications.  The scheme makes use of BACKscattered GAMMas On Nucleons  (BACKGAMMON).  Current accelerator infrastructures could be utilized to produce the muon beams, such as the Electron-Ion Collider (EIC) at Brookhaven National Laboratory in the United States.  We discuss the implementation of BACKGAMMON at the EIC. 

\end{abstract}

\noindent {\bf Introduction}\\
 There has been growing interest in high intensity muon beams for a variety of applications, such as a muon-muon collider \cite{Muon collider}, or perhaps a muon-ion collider at Brookhaven National Laboratory in the United States \cite{MuI collider, MuI collider Ethan}.  One approach being studied uses a proton driver, which employs an intense proton beam that impinges upon a target, such as graphite, thereby producing pions, which decay into $\sim 10^{14}$ muons/s \cite{Muon rate from protons}.  An alternative approach, called the {\em Gamma Factory}, uses high energy photons resulting from resonant backscattering of laser photons off atomic beams of partially stripped ions circulating in CERN's Large Hadron Collider storage rings \cite{GF}.  The high energy photons in turn strike a graphite target to produce pions whose decay should also be able to achieve $\sim 10^{14}$ muons/s.  Yet another approach, called {\em BACKscattered GAMMas On Nucleons (BACKGAMMON)} \cite{Backgammon}, and which we present here, uses high energy photons, which are derived from Compton backscattered laser pulses from an intense electron beam, that similarly impinge upon a stationary target to produce pions, which decay into $\sim 10^{14}$~muons/s.   In the latter approach, it may be possible to utilize an existing infrastructure as the electron source, such as the electron storage ring at Brookhaven National Laboratory's Electron-Ion Collider in the United States.  This could be the next phase of research at that facility.  As for BACKGAMMON, it was proposed some time ago as a scheme for scattering ultra high energy backscattered photons on targets for a variety of studies, including CP violation in B-meson decay \cite{Backgammon}, polarized $\tau$ decay \cite{tau decay}, and various other applications to heavy flavor physics \cite{Heavy flavor physics}.

\vskip.20in

\noindent {\bf Backscattered Photon Energy}\\
The analysis of the electron-photon scattering is considered in the laboratory frame,
wherein both the electron and low energy laser photon beams are propagating in opposite directions.  For an excellent discussion, see Ref.~\cite{Chaikovska}.  The general case is depicted in Fig.~1, where $p_1, k_1, k_2$ are the 4-momenta of the incident electron, incident photon, and outgoing scattered photon, respective; and $\phi_1$ is the angle between the incident directions of the electron and photon beams, $\phi_2$ is the angle between the incident electron and scattered photon beams, and $\theta$ is the angle between the incident and scattered photon directions.

\vskip  0.2in
\begin{figure}[H]
\centering
 \includegraphics[height=7cm]{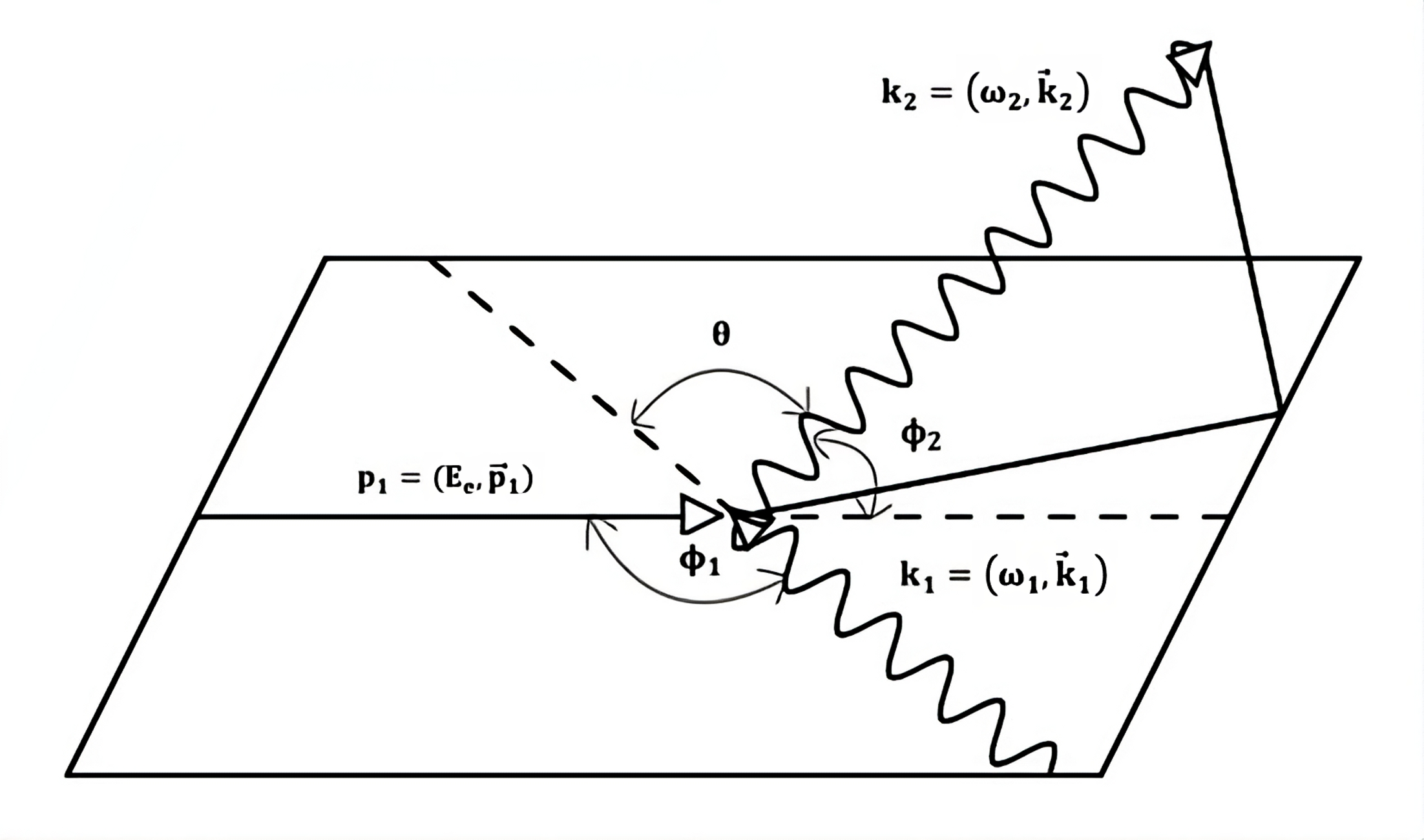}
\caption*{Fig. 1. Schematic of Compton Backscattering} \label{fig1}
\end{figure}

\vskip  0.2in

For our case, the photon beam is backscattered with $\theta \sim \pi$, and the resulting high energy backscattered photon beam is confined within a cone of opening
angle $\sim \dfrac{1}{\gamma}$ relative to the electron’s incident direction of motion, with $\gamma$ being the usual
Lorentz contraction factor for the electron beam.  We take the incident electron to have energy $E_e$ and incident photon to have energy $\omega_1$, with the electron rest energy being $E_0$.  The scattered photon energy $\omega_2$ is given by

\begin{equation} \label{eq:omega}
\omega_2 = \frac{\omega_1 (1-\beta \cos \phi_1)}{1-\beta \cos \phi_2 + \dfrac{\omega_1}{E_e} (1-\cos \theta)}\, \, ,
\end{equation}
\\
\noindent where as usual, $\beta=\dfrac{v}{c}$;   When $\omega_1 < < E_e$, as in our case, we have the approximation

\begin{equation} \label{eq:omega2approx}
\omega_2 \approx \frac{4 \gamma^2 \omega_1}{1+4\gamma^{2} \dfrac{\omega_1}{E_e}} \, \, .
\end{equation}

\vskip.20in

\noindent {\bf Unpolarized Total Cross Section}\\
Following Landau, Lifshitz, Berestetskii and Pitaevskii  \cite{Landau}, we set
\begin{equation}  \label{eq:x1}
x_1 = 2\gamma \frac{\omega_1}{E_0} (1-\beta \cos \phi_1).
\end{equation}

\noindent The total unpolarized Compton cross section $\sigma_C^{unp}$ is then
\begin{equation}  \label{eq:sigmaunp}
\sigma_C^{unp} = 2 \pi r_{0}^{2} \frac{1}{x_1} [(1-\frac{4}{x_1} - \frac{8}{x_{1}^{2}}) \ln (1+x_1) + \frac{1}{2} + \frac{8}{x_1}
    - \frac{1}{2(1+x_1)^2)}],
\end{equation}

\noindent where $r_0  = 2.81794 \times 10^{-15}$ m is the classical electron radius.

\vskip.20in

\noindent {\bf Total Cross Section with Polarization Included}\\
To include polarization effects, one write can write \cite{Ginzburg}
\begin{equation}  \label{eq:Cross section total}
\sigma_C = \sigma_C^{unp} + 2 \lambda_e P_c \sigma_1,
\end{equation} 

\noindent where $\lambda_e$ is the mean electron helicity $(|\lambda_e| \leq \dfrac{1}{2})$, $P_c$ is the Stokes parameter that gives
the degree of photon circular polarization given by
\begin{equation}  \label{eq:Stokes}
P_c = \frac{(I_L - I_R)}{I_0},
\end{equation}

\noindent with $I_L$, $I_R$, and $I_0$ being the left circularly polarized, right circularly polarized, and total photon intensities.  Finally, 
\begin{equation} \label{eq:Polarized cross section}
\sigma_1 = \frac{2 \sigma_0}{x_1} [(1 + \frac{2}{x_1}) \log (1+x_1) - \frac{5}{2} + \frac{1}{1+x_1} - \frac{1}{2(1+x_1)^2}]
\end{equation}

\noindent with $\sigma_0 = \pi (\dfrac{e^2}{m_e c^2}) = 2.5 \times 10^{-29}\, \mbox{m}^2$.

\vskip.20in

\noindent {\bf Luminosity of Electron-Photon Collisions}\\
We use the Suzuki formula \cite{Suzuki}, where the luminosity
\begin{equation} \label{eq:Suzuki}
L = N_e N_{ph} f \frac{\cos (\dfrac{\phi}{2})}{2 \pi} \frac{1}{\sqrt{(\sigma^{2}_{ye} +  \sigma^{2}_{yph}} \sqrt{(\sigma^{2}_{xe} +  \sigma^{2}_{xph})
\cos^{2} (\dfrac{\phi}{2}) +  (\sigma^{2}_{ze} +  \sigma^{2}_{zph} ) \sin^{2} (\dfrac{\phi}{2} ) }},
\end{equation}
\\
\noindent where $N_e$ is the number of electrons in a bunch, $N_{ph}$ is the numbers of photons in a laser pulse, $f$ is the electron-photon bunch collision frequency, $\phi$ is the crossing angle of the lines defined by the incident electron and photon momenta, and the $\sigma$'s define the rms sizes of electron and photon beam dimensions at the interaction point, with $z$ being the longitudinal direction according to Suzuki’s notation. The electron rms beam sizes are given by the Electron-Ion Collider (EIC) Design parameters \cite{EIC}, the laser pulse transverse sizes are taken to correspond to those of the electron bunches, while the laser pulse length is defined by the laser system discussed below.

From the luminosity $L$, the rate $R$ of backscattered photons is given by
\begin{equation}
R\equiv \frac{dN}{dt} = \sigma_C L,
\end{equation}

\noindent where $N$ is the number of backscattered photons.

\vskip.20in
\noindent {\bf Laser Parameters}\\
The frequency $f$ at which electron bunches come into the interaction point with the laser pulses is equal to the number of electron bunches in the storage ring (1,160) times the revolution frequency of the electrons around the circumference (3,833.94 meters), or

\begin{equation}
f = \beta c \frac{1160}{3833.94} = 90.7 \, \, \mbox{MHz} \, .
\end{equation}

M{\"u}ller {\em et al.} have reported developing a laser system centered at 1046 nm wavelength $\lambda$, 254 fs pulse duration, 80 MHz pulse repetition rate, and 10.4 kW average output power P \cite{Muller}.  We note that this is not quite at the 90.7 MHz repetition rate of the electron bunches at the EIC; however, they are close enough that we take the repetition rate of both the electron bunches and laser pulses to be 90.7 MHz.  For this laser system, the energy per laser pulse is 114.657 $\mu$J and the number of incident photons per pulse is $6.03 \times 10^{14}$.  The authors report that the average power can be pushed to 100 kW, which would increase the number of photons per pulse by an order of magnitude.

\vskip.20in
\noindent {\bf Rate of Backscattered Photons}\\
Table~1 contains updated EIC parameters and the laser system modified to 90.7 MHz pulse rep rate.

\newpage

\begin{center}
Table 1. Input Parameters    \vskip0.1in
\begin{tabular}{|c|c|c|c|c|c|c|c|}                                
 \hline
  $N_e$                           & $N_{ph}$                         &    $E_e$             &  $\omega_1$  &  $\phi_1$  &  $\phi_2$   &  $\theta$  & $\phi$       \\    \hline
  $1.72 \times 10^{11}$   &   $6.03 \times 10^{14}$  &  $4.5$ GeV        & $1.19$ eV      &   $\pi$        &   $0$         & $\pi$         &  $0$   \\ \hline                             
 \hline
  $\lambda_e$    & $P_c$     &    $\epsilon_x$     &  $\epsilon_y$  &  $\beta^{\ast}_x$  &  $\beta^{\ast}_y$   &  $f$    &   \\    \hline
    $\frac{1}{2}$   &   $1$        &  $20$ nm            & $1.2$ nm         &   $1$ cm              &   $1$ cm               & $90.7$ MHz  &      \\ \hline 
\end{tabular}
\end{center}

\vskip.10in

The number of photons per laser pulse $N_{ph}$ is calculated from the laser system as
\begin{equation}
N_{ph} = \dfrac{(\dfrac{P}{f})}{(\dfrac{hc}{\lambda})} \, \, .
\end{equation}

\noindent The laser is focused so that its transverse rms pulse sizes $\sigma_{xph}$ and $\sigma_{yph}$ match those of the electron bunches, which are
\begin{subequations}
\begin{align}
\sigma_{xe}  &=  \sqrt{\epsilon_x \beta^{\ast}_x} \label{eq:a}  \\
\sigma_{ye}  &= \sqrt{\epsilon_y \beta^{\ast}_y}  \label{eq:b} \,  \, ,
\end{align}
\end{subequations}

\noindent with $\epsilon_x$, $\epsilon_y$ being the electron beam emittances and  $\beta^{\ast}_x$,  $\beta^{\ast}_y$ the electron accelerator lattice functions at the interaction point with the laser beam, giving the values shown in Table~2.

\vskip  0.1in
\begin{center}
Table 2. Electron and Laser Beam Sizes     \vskip0.1in
\begin{tabular}{|c|c|c|c|c|c|}                                
 \hline
  $\sigma_{xe}$           & $\sigma_{xph}$                &    $\sigma_{ye}$              &   $\sigma_{yph}$             &  $\sigma_{ze}$             &   $\sigma_{zph}$         \\    \hline
  $14.14 \, \mu\mbox{m}$   &   $14.14 \, \mu\mbox{m}$   &  $3.46\, \mu\mbox{m}$       & $3.46 \, \mu\mbox{m}$      &   $11.0\,  \mbox{mm}$        &   $76.1 \, \mu\mbox{m}$       \\ \hline                             
\end{tabular}
\end{center}

\vskip.10in

\noindent where $\sigma_{ze}$ is taken from updated EIC parameters, and the photon pulse length is given by $\sigma_{zph} = c \times 254$ fs.

Finally, we obtain Table 3 for the derived backscattered photon energy  $\omega_2$, total Compton cross section $\sigma_C $, luminosity $L$, and production rate $R$ of backscattered photons for $E_e = 4.5$~GeV.

\vskip  0.1in
\begin{center}
Table 3. Derived Quantities  ($E_e = 4.5$ GeV)     \vskip0.1in
\begin{tabular}{|c|c|c|c|}                                
 \hline
  $\omega_2$          & $\sigma_C $                                  &    $L$                                                     &   $R$     \\    \hline
  $340$ MeV            &   $6.05\times 10^{-25}$ cm$^2$   &  $1.53 \times 10^{39}/$cm$^{2}/$s        & $0.93 \times10^{15}/$s      \\ \hline                             
\end{tabular}
\end{center}

\vskip.20in

\noindent {\bf Muon Production}\\
Tables 1-3 apply essentially to the current state of the electron and laser systems.  The rate of muon production from the current systems is a couple of orders of magnitude less than the goal of $\sim 10^{14}$ muons/s.  However, there are several possibilities of enhancing the input parameters.  First, Ref.~\cite{Muller} states that it is possible to increase tthe laser average power by an order of magnitude up to 100 kW.  Also, further research is needed to achieve another order of magnitude from the number of electrons in a bunch $N_e$, accelerator lattice parameters $\beta_x, \beta_y$, and/or electron beam emittances $\epsilon_x, \epsilon_y$.  Thus, we can increase the rate $R$ of backscattered photon production by two orders of magnitude to $10^{17}$/s.  In terms of implementing BACKGAMMON at the EIC, there is at least a decade of time to research these issues before it will be possible there, due to the new EIC just starting its research program. 

Considering the above comments, to produce the muon beam, $10^{17}$/s backscattered photons of energy 340 MeV impinge upon a stationary target, which we take to be graphite.  After varying the target length and radius, we choose the length to be 30~cm and radius 2.5~cm.  Both pions and muons are produced, where the pions subequently decay into muons.  The production of muons through the pion channel yields several orders of magnitude more muons than the direct muon channel.  A major advantage of the pion channel is that a 340 MeV backscattered photon scattering off a stationary proton (neutron) yields an invariant mass of 1232 MeV (1234 MeV), which sits right at the $\Delta(1232)$ resonance, thereby boosting the photo-absorption cross section of the process \cite{GF}.  An added benefit of producing muon beams using the BACKGAMMON scheme is that the numbers of positive and negative muons are comparable. This should be a tremendous advantage for a $\mu^{+}$-$\mu^{-}$ collider.

\vskip.20in

\noindent {\bf Pion Capture}\\
For energies E $\sim$ 340~MeV, the main process leading to $\gamma$ beam absorption in matter is the formation of electron-positron pairs.  The electromagnetic cascade is formed inside the target, producing secondary photons, electrons, and positrons.  Only the small fraction of the photons produce pions.  The production and collection of pions and their decay product muons must be optimized in order to produce and collect large numbers of muons.

The method of producing muons used in this paper closely follows the proposed scheme of the Gamma-Factory in Ref.~\cite{GF}.  The Gamma Factory uses a primary photon beam of energy up to 300~MeV, while here we use a photon beam of energy 340~MeV.  Following the approach taken in Ref.~\cite{GF}, we use two methods of selecting pions against the dominant contributions of electrons and positrons.  The first method uses the fact that the electrons and positrons have velocities close to the speed of light ($\beta _e \approx 1$), while the pions, carrying kinetic energy comparable to their mass, are non-relativistic ($\beta _{\pi} \ll 1$).  The second method exploits the differences in their production mechanisms, leading to their  distinct kinematical characteristics. Particularly, the pions are produced with a larger angle with respect to the primary photon beam compared to the electrons and positrons~\cite{GF}. Thus, the particles having transverse momenta higher than $30\,$MeV/c and collected in the angular region $\theta \geq  40^{\circ}$ are predominantly pions. 

\vskip.20in

\noindent {\bf Simulations}\\
Simulations have been performed using GEANT4.11.2.  Comparisons were made with the earlier version, GEANT4.10.6, that showed good consistency between the two versions.  In Fig.~ 2, transverse momentum (2a), total momentum (2b), and kinetic energy (2c) distributions of positive and negative pions 1 mm from the end of the target are shown for the optimal target material, which is graphite~\cite{GF}.  The optimised target length is $30\,$cm with a radius of $2.5\,$cm, within the selected pion collection angular region of $40^{\circ}\,$--$\,120^{\circ}$ and transverse momentum greater than $30\,$MeV/c.  For a photon rate of $10^{17}$/s and beam energy $340\,$MeV, one can achieve pion fluxes greater than $10^{13}$/s for both positive and negative pions. The pions are produced over a narrow range of momentum and transverse-momentum distributions. The widths of the transverse and total momentum distributions are, respectively, $\sigma _{p_T} = 50\,$MeV/c and $\sigma_{p_{tot}} = 27\,$MeV/c.

\begin{figure}[H]
\centering
\begin{subfigure}{.5\textwidth}
  \centering
  \includegraphics[height=5.6cm]{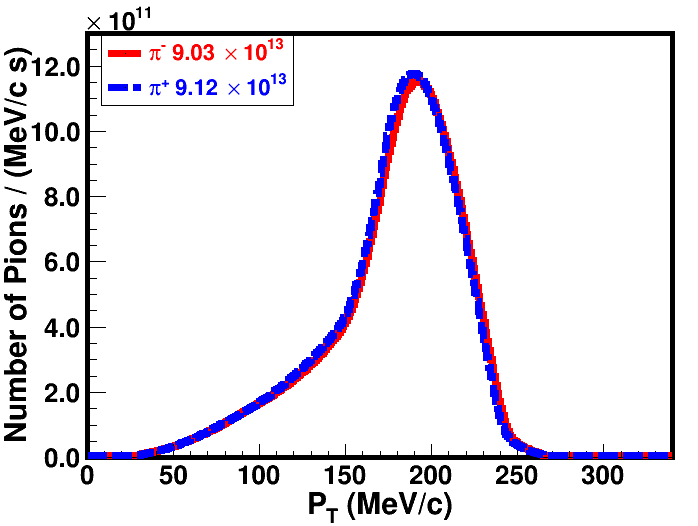}
  \caption*{a. Pion Transverse Momentum Distribution} \label{fig2a}
  \end{subfigure}%
\begin{subfigure}{.5\textwidth}
  \centering
 \includegraphics[height=5.6cm]{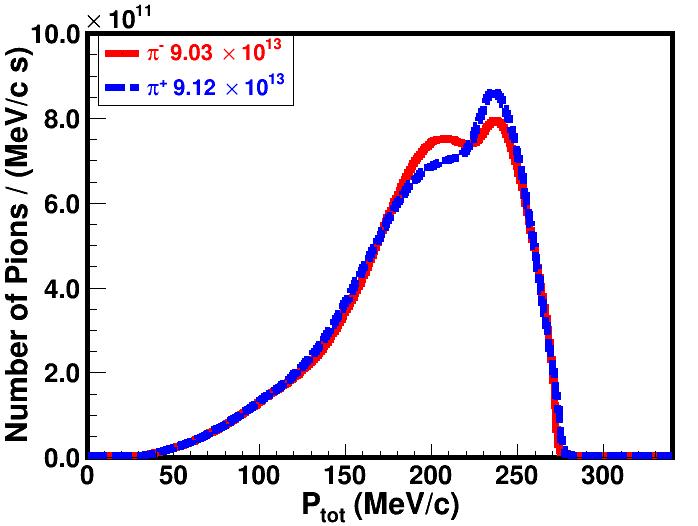}
\caption*{b. Pion Total Momentum Distribution} \label{fig2b}
\end{subfigure}%
\vskip 0.2in 
\begin{subfigure}{.5\textwidth}
  \centering
 \includegraphics[height=5.6cm]{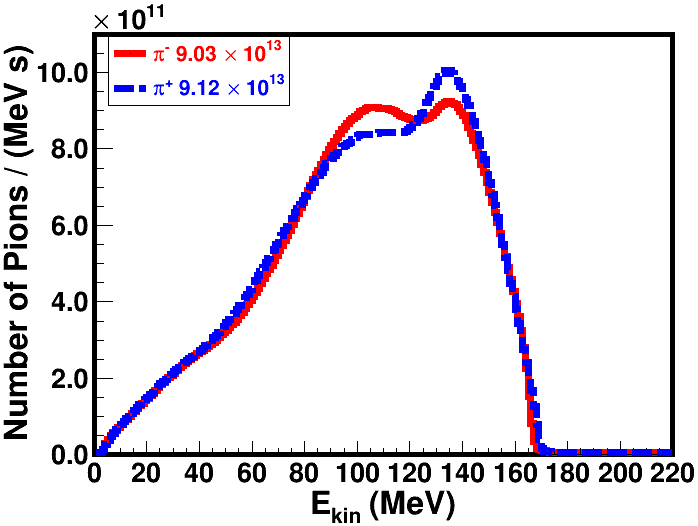}
\caption*{c. Pion Kinetic Energy Distribution} \label{fig2c}
\end{subfigure}
\caption*{Fig. 2. Pion Kinematic Distributions 1 mm from the end of the graphite target \\({\footnotesize Photon energy=340 MeV, Photon rate=$10^{17}$/s, Target length=30 cm, Target radius=2.5 cm, \\pion $p_T\geq 30\,$MeV/c, pion collection angular region is $40^{\circ}\,$--$\,120^{\circ}$)}}
\label{fig:Kinetics Pion}
\end{figure}

\vskip  0.2in
\begin{figure}[H]
\centering
\begin{subfigure}{.5\textwidth}
  \centering
  \includegraphics[height=5.6cm]{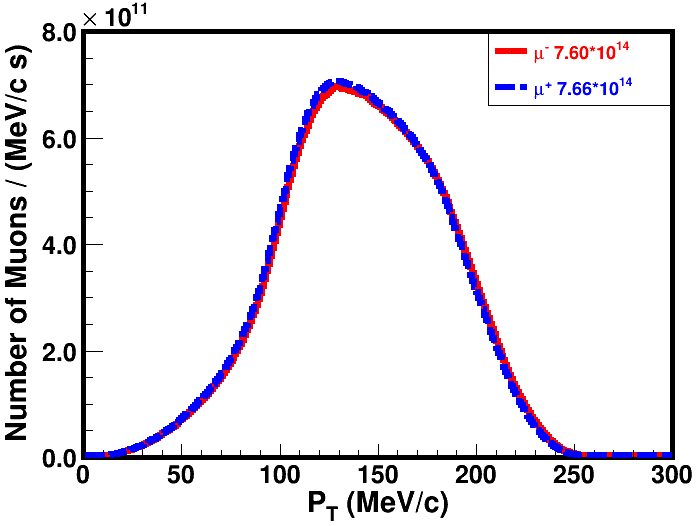}
  \caption*{a. Muon Transverse Momentum Distribution} \label{fig3a}
  \end{subfigure}%
\begin{subfigure}{.5\textwidth}
  \centering
 \includegraphics[height=5.6cm]{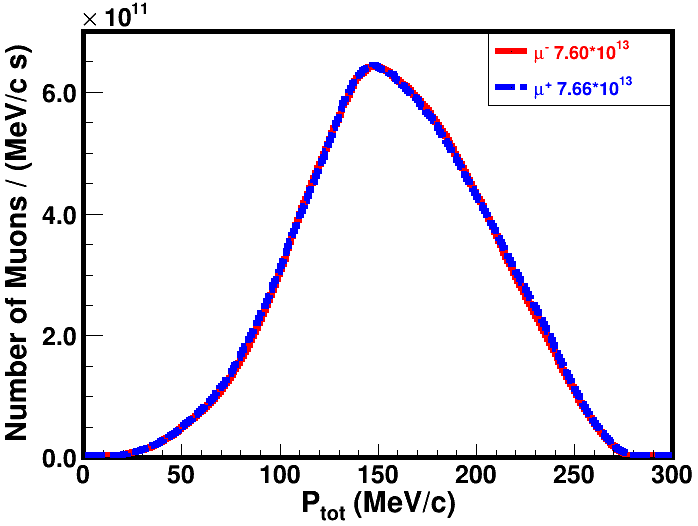}
\caption*{b. Muon Total Momentum Distribution} \label{fig3b}
\end{subfigure}%
\vskip 0.1in 
\begin{subfigure}{.5\textwidth}
  \centering
 \includegraphics[height=5.6cm]{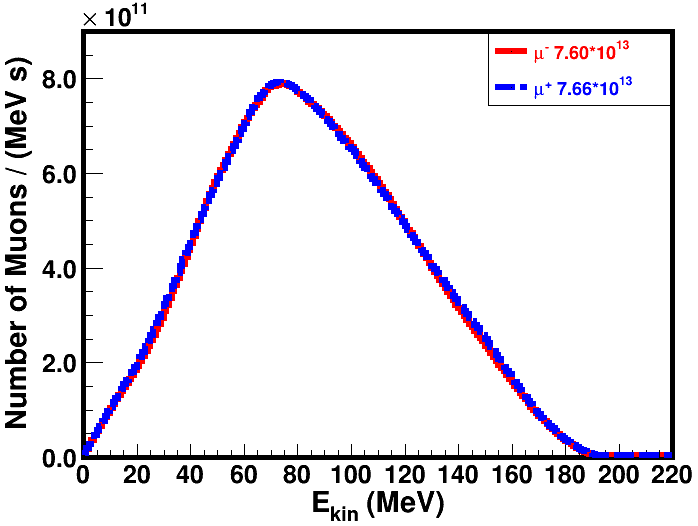}
\caption*{c. Muon Kinetic Energy Distribution} \label{fig3c}
\end{subfigure}
\caption*{Fig. 3. Muon Kinematic Distributions 30~m from the center of the graphite target \\({\footnotesize Photon energy=340 MeV, Photon rate=$10^{17}$/s, Target length=30cm, Target radius=2.5 cm,  \\muons resulting from pions decaying before reaching 30~m from center of target, pion \\$p_T\geq 30\,$MeV/c, pion collection angular region is $40^{\circ}\,$--$\,120^{\circ}$)}}
\label{fig:Kinetics}
\end{figure}

\vskip  0.1in
\noindent Fig.~3  shows the transverse momentum, total momentum, and kinetic energy distributions of muons coming from pions decaying before reaching $30\,$m from the center of the target.  The plots are made assuming that all the initial pions satisfy the pion selection criteria specified above.

Finally, Fig.~4 gives the rate of muon production vs. pion propagation length from the center of the graphite target, yielding $\sim 10^{14}$ muons/s, which was the goal.

\vskip  0.2in
\begin{figure}[H]
\centering
 \includegraphics[height=7cm]{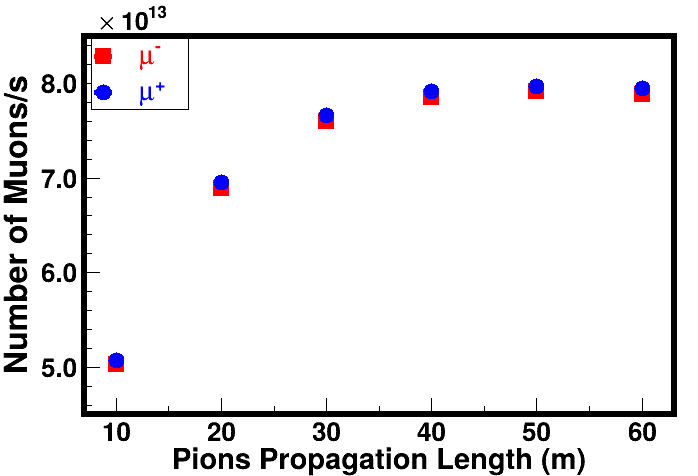}
\caption*{Fig. 4. Rate of Muon Production vs. Pion Propagation Length from the center of the graphite target} \label{fig4}
\end{figure}

\vskip  0.1in

The presented studies can be considered as an initial, exploratory step.  Conceptual and technical designs of the pion and muon collection scheme needs to be developed.  Such studies have been performed for the proton beam-driven muon sources, as outlined in Ref.~\cite{Palmer}.  The idea can also be implemented for a photon beam-driven muon source.  An ultra high-field solenoid can extend the length of a target upon which a photon beam impinges.  This target solenoid collects pions with a large momentum spread and large angles, and guides them downstream into a long solenoid channel, where they decay to muons.  Quantitatively, the effect of the solenoidal field profile on the capture efficiency can be evaluated by calculating the muon yield, which is defined as the number of particles that fall within a reference acceptance.  A dedicated study of the pion capture system is required.

To conclude, we have presented a scheme called BACKGAMMON, which has the possibility to produce $\sim 10^{14}$ muons/s for muon-muon and muon-ion colliders, as well as for other applications.  One big challenge is the circulating electron beam lifetime, which is on the order of 20 to 200 ms.  However, with future research innovations, this problem can be overcome.  An example of a revolutionary method of replenishing circulating beams is the {\em Swap-Out} mode that was invented at the Advanced Photon Source at Argonne National Laboratory in the United States \cite{Swap-out}.  It uses fast kicker magnets to extract and dump the stored electron bunches while replacing them with fresh bunches from the injector, all within a few nanoseconds. This process maintains the beam current at a constant value.  Indeed, there is considerable research that needs to be pursued to make BACKGAMMON a reality; however, there is a decade or more before Brookhaven National Laboratory can even begin to implement it.

 \vskip.20in

\newpage
\noindent {\bf Acknowledgements}\\
\noindent  The authors would like to thank the following persons for their many insightful comments: J. Scott Berg, Ethan Cline, Abhay Deshpande, Mieczyslaw Witold Krasny, Christoph Montag, and Vadim Sajaev.  The study was partly funded by National Science Foundation Grant No.~NSF PHY-2310078 at Michigan State University's Facility for Rare Isotope Beams.

The authors of this paper were ordered alphabetically.

\end{sf}

\end{document}